\documentclass[twocolumn,trackchanges,floatfix]{aastex7}
\usepackage{mathtools} 
\usepackage{csvsimple}
\usepackage{longtable}
\usepackage{footmisc}
\usepackage{mhchem}
\renewcommand{\thefootnote}{\arabic{footnote}}

\shortauthors{Voitiuk et al.}
\graphicspath{{./}{figures/}}
\usepackage{orcidlink}

\begin{document}

\title{ZTF J2229+1928: A 23 Minute Eclipsing Double White Dwarf Binary}
\altaffiliation{This paper includes data gathered with the 6.5 meter Magellan \\ Telescopes located at Las Campanas Observatory, Chile.}

\author[0009-0000-8838-1618]{Viktoriia Voitiuk}
\affiliation{Department of Physics, Massachusetts Institute of Technology, Cambridge, MA 02139, USA}
\affiliation{Kavli Institute for Astrophysics and Space Research, Massachusetts Institute of Technology, Cambridge, MA 02139, USA}
\email{voit2206@mit.edu}

\author[0000-0002-5812-3236]{Aaron Householder}
\altaffiliation{NSF Graduate Research Fellow}
\affiliation{Department of Earth, Atmospheric and Planetary Sciences, Massachusetts Institute of Technology, Cambridge, MA 02139, USA}
\affiliation{Kavli Institute for Astrophysics and Space Research, Massachusetts Institute of Technology, Cambridge, MA 02139, USA}
\email{aaron593@mit.edu}

\author[0000-0002-7226-836X]{Kevin B. Burdge}
\affiliation{Department of Physics, Massachusetts Institute of Technology, Cambridge, MA 02139, USA}
\affiliation{Kavli Institute for Astrophysics and Space Research, Massachusetts Institute of Technology, Cambridge, MA 02139, USA}
\email{kburdge@mit.edu}

\author[0000-0002-7191-4403]{Christopher Layden}
\affiliation{Department of Physics, Massachusetts Institute of Technology, Cambridge, MA 02139, USA}
\affiliation{Kavli Institute for Astrophysics and Space Research, Massachusetts Institute of Technology, Cambridge, MA 02139, USA}
\email{clayden7@mit.edu}

\author[0000-0002-0568-6000]{Joheen Chakraborty}
\affiliation{Department of Physics, Massachusetts Institute of Technology, Cambridge, MA 02139, USA}
\affiliation{Kavli Institute for Astrophysics and Space Research, Massachusetts Institute of Technology, Cambridge, MA 02139, USA}
\email{joheen@mit.edu}

\author[0000-0003-4780-4105]{Emma T. Chickles}
\affiliation{Department of Physics, Massachusetts Institute of Technology, Cambridge, MA 02139, USA}
\affiliation{Kavli Institute for Astrophysics and Space Research, Massachusetts Institute of Technology, Cambridge, MA 02139, USA}
\email{echickle@mit.edu}

\author[0000-0002-6401-778X]{Jack Dinsmore}
\affiliation{Department of Physics, Stanford University, Stanford, CA 94305, USA}
\affiliation{Kavli Institute for Particle Astrophysics and Cosmology, Stanford University, Stanford, CA 94305, USA}
\email{jtd@stanford.edu}

\author[0000-0001-8467-9767]{Gabor Furesz}
\affiliation{Kavli Institute for Astrophysics and Space Research, Massachusetts Institute of Technology, Cambridge, MA 02139, USA}
\email{gfuresz@mit.edu}

\author[0000-0003-1361-985X]{Juliana García-Mejía}
\altaffiliation{51 Pegasi B Fellow, MIT Pappalardo Physics Fellow}
\affiliation{Kavli Institute for Astrophysics and Space Research, Massachusetts Institute of Technology, Cambridge, MA 02139, USA}
\affiliation{Department of Physics, Massachusetts Institute of Technology, Cambridge, MA 02139, USA}
\affiliation{Center for Astrophysics \textbar\ Harvard \& Smithsonian, 60 Garden Street, Cambridge, MA 02138, USA}
\email{jgarciam@mit.edu}

\author[0000-0002-3239-5989]{Aurora Y. Kesseli}
\affiliation{IPAC, Mail Code 100-22, Caltech, 1200 E. California Boulevard, Pasadena, CA 91125, USA}
\email{aurorak@ipac.caltech.edu}

\author[0000-0001-6331-112X]{Geoffrey Mo}
\email{gmo@caltech.edu}
\affiliation{Division of Physics, Mathematics, and Astronomy, California Institute of Technology, Pasadena, CA 91125, USA}
\affiliation{The Observatories of the Carnegie Institution for Science, Pasadena, CA 91101, USA}

\author[0000-0003-0412-9664]{David Osip}
\email{dosip@carnegiescience.edu}
\affiliation{Las Campanas Observatory, Carnegie Institution for Science, Colina el Pino, Casilla 601 La Serena, Chile}

\author[0000-0001-7246-5438]{Andrew Vanderburg}
\affiliation{Center for Astrophysics \textbar\ Harvard \& Smithsonian, 60 Garden Street, Cambridge, MA 02138, USA}
\email{avanderburg@cfa.harvard.edu}

\author[0000-0003-0097-4414]{Yapeng Zhang}
\altaffiliation{51 Pegasi b Fellow}
\affiliation{Department of Astronomy, California Institute of Technology, Pasadena, CA 91125, USA}
\email{yapzhang@caltech.edu}

\begin{abstract}
We report the discovery of ZTF J2229+1928, a detached and eclipsing DA double white dwarf binary in a 23-minute orbit. The system was identified in a GPU-accelerated search of ZTF photometry and confirmed through follow-up observations with the high speed imager proto-Lightspeed on the 6.5-meter Magellan Clay telescope. Combining the proto-Lightspeed eclipse model with mass-radius relations gives component radii and masses of $R_1 = 0.0146^{+0.0009}_{-0.0008} \, R_\odot$, $R_2 = 0.0186^{+0.0010}_{-0.0008} \, R_\odot$, and $M_1 = 0.49^{+0.08}_{-0.07} \, M_\odot$, $M_2 = 0.37^{+0.06}_{-0.05} \, M_\odot$. Fitting the Keck/LRIS spectra gives component temperatures of approximately 11{,}100\,K and 8{,}200\,K, and scaling this model to the broadband spectral energy distribution gives a distance of $\sim$965\,pc. Its short orbital period places it firmly in the millihertz gravitational-wave regime, and we estimate an expected signal-to-noise ratio $\mathrm{SNR} \approx 5.3$ with the upcoming \textit{LISA} mission over its nominal four-year duration. ZTF J2229+1928 thus adds to the small but growing sample of eclipsing ultracompact binaries that will serve as important sources for future space-based gravitational-wave observatories.
\end{abstract}

\section{Introduction}\label{sec:intro}

\begin{figure*}[!t]
    \centering
    \includegraphics[width=\linewidth]{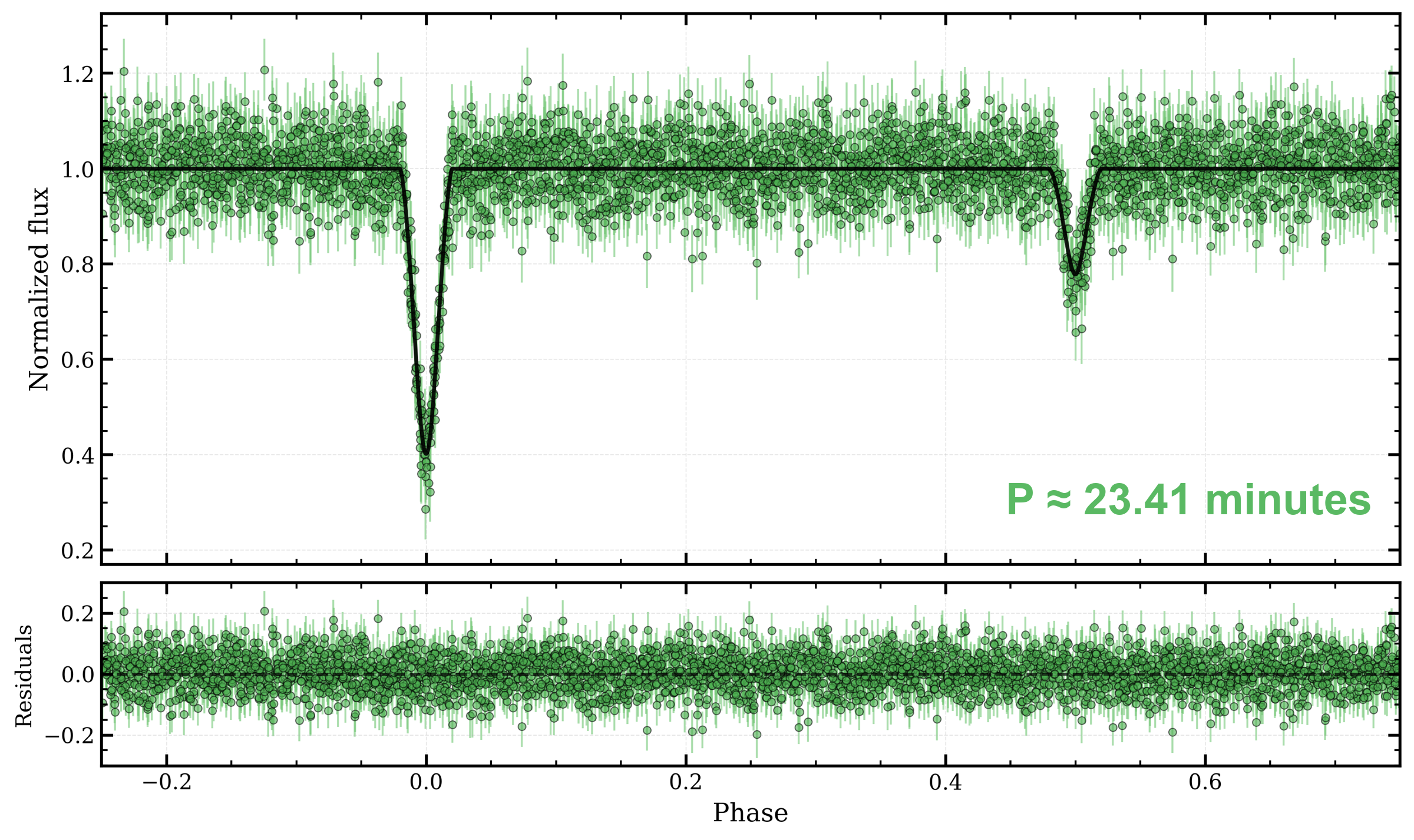}
    \includegraphics[width=\linewidth]{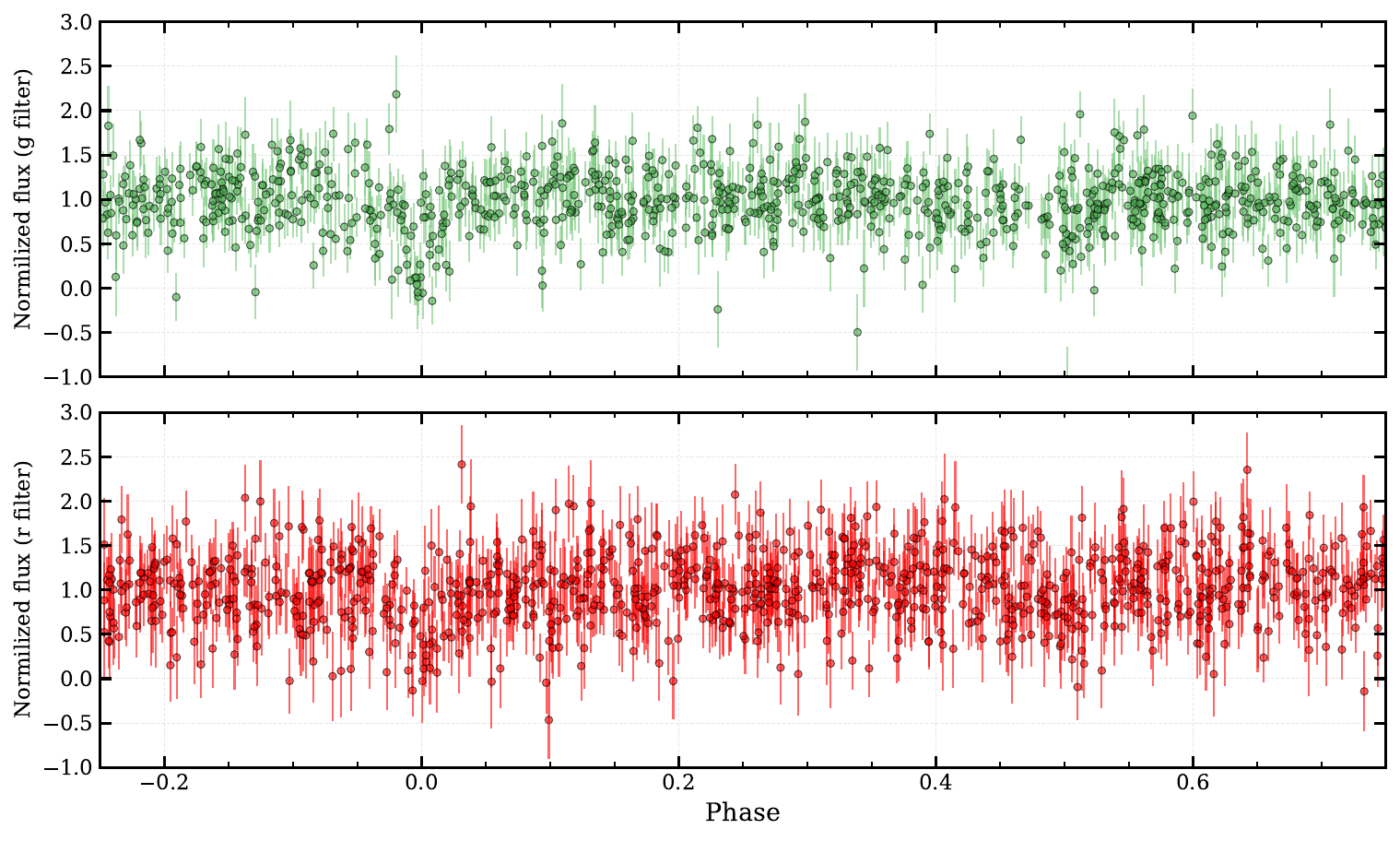}
    \caption{Top) High-speed follow-up light curve of ZTF J2229+1928 (green) obtained with proto-Lightspeed on Magellan \citep{Layden2026,Layden2026b}, phase-folded to the period recovered with the ZTF search ($P \approx 23.41$ minutes). The best fit \texttt{ellc} model is plotted in black. Bottom) ZTF g-band (top green) and r-band (bottom red) light curves of ZTF J2229+1928 used to discover the object, phase-folded to the same period.}
    \label{fig:proto-Lightspeed}
\end{figure*}

The growing sample of compact double white dwarf (DWD) binaries has transformed our ability to study both the outcomes of stellar binary evolution and the progenitors of gravitational-wave sources detectable by the upcoming \textit{Laser Interferometer Space Antenna (LISA)}. Over the past decade, large spectroscopic efforts such as the ELM Survey and SPY have identified more than a hundred close double white dwarf binaries \citep[e.g.,][]{Brown2020,Napiwotzki2020}. More recently, wide-field time-domain surveys have expanded the sample to include several eclipsing systems, which provide valuable additional constraints on the properties of both components in the binary. In particular, the Zwicky Transient Facility (ZTF; \citealt{Bellm2019,Graham2019,Masci2019}) has discovered several short-period detached eclipsing DWDs \citep{Burdge2019,Burdge2020a,Burdge2020b,Coughlin2020,vanRoestel2025}, as well as eclipsing AM CVn systems and other accreting ultracompact binaries whose donors may be white dwarfs \citep{vanRoestel2022,Chakraborty2024}. 

In this paper, we report the discovery of ZTF J2229+1928, a detached and eclipsing DWD with an orbital period of just 23.4 minutes. We describe the discovery of ZTF J2229+1928 from ZTF photometry (Section \ref{sec:discovery}), present high-speed follow-up with proto-Lightspeed on Magellan (Section \ref{sec:followup}), and model the high-speed imaging light curves, Keck/LRIS spectra, and spectral energy distribution to measure the parameters of the system (Section \ref{sec:modeling}). We then compute the expected gravitational-wave signal with \textit{LISA} and discuss the orbital decay and fate of the system (Section \ref{sec:discussion}). Finally, in Section \ref{sec:conclusion}, we summarize our findings and outline prospects for future characterization of ZTF J2229+1928. 

\section{Discovery of ZTF J2229+1928}\label{sec:discovery}
\subsection{Detection in ZTF}
ZTF J2229+1928 was identified as part of a systematic search for short-period eclipsing systems in photometric data from the Zwicky Transient Facility (ZTF), which surveys the northern sky with 30\,s exposures. We analyzed light curves for 1.3 billion Pan-STARRS sources using a GPU-accelerated implementation of the Box-Least Squares (BLS) algorithm \citep{Kovacs2002}. Our search flagged ZTF J2229+1928 as one of the most significant short-period candidates, with a strong BLS peak near $P \approx 23.41$ minutes. The period reported in Table \ref{tab:params}, $P = 23.413090 \pm 0.000013$ minutes, is the location of this BLS peak in the ZTF light curve, with the uncertainty given by the half width at half maximum of the periodogram peak. Unlike many known DWDs, ZTF J2229+1928 is so faint that \textit{Gaia} does not even detect it as a point source, so it is not listed in published white dwarf catalogs. Nevertheless, because our search was performed over 1.3 billion Pan-STARRS sources rather than a catalog-selected subsample, the system was successfully recovered.  

\subsection{Confirmation with proto-Lightspeed}\label{sec:followup}
On UT 2025 September 15, we obtained high-speed unfiltered photometric follow-up of ZTF J2229+1928 with proto-Lightspeed during a commissioning run on the Magellan Clay telescope \citep{Layden2026}. The observations ran continuously for approximately 4.7 hours, yielding 3350 consecutive 5-second exposures that captured multiple consecutive eclipses. proto-Lightspeed records each observation as a merged data cube, in which thousands of short-exposure frames are stacked with GPS-synchronized timestamps. We extracted the individual frames and performed aperture photometry on ZTF J2229+1928 and nearby comparison stars using a custom photometry pipeline that has also been used in recent work (e.g., \citealt{Chickles2026}). This pipeline includes the nonlinearity calibration presented in \citet{Layden2026}, which corrects the flux measured in each pixel to account for photogenerated charge that is trapped during readout. A differential light curve was constructed by dividing the target flux by the combined flux of stable comparison stars to correct for transparency variations. We also rejected outliers by removing data points with flux uncertainties larger than the median uncertainty plus $3\sigma$, which removed 73 of the 3350 data points. The final differential light curve clearly shows repeated narrow eclipses at the period recovered in ZTF (Figure \ref{fig:proto-Lightspeed}). 

\begin{figure*}[!t]
    \centering
    \includegraphics[width=\linewidth]{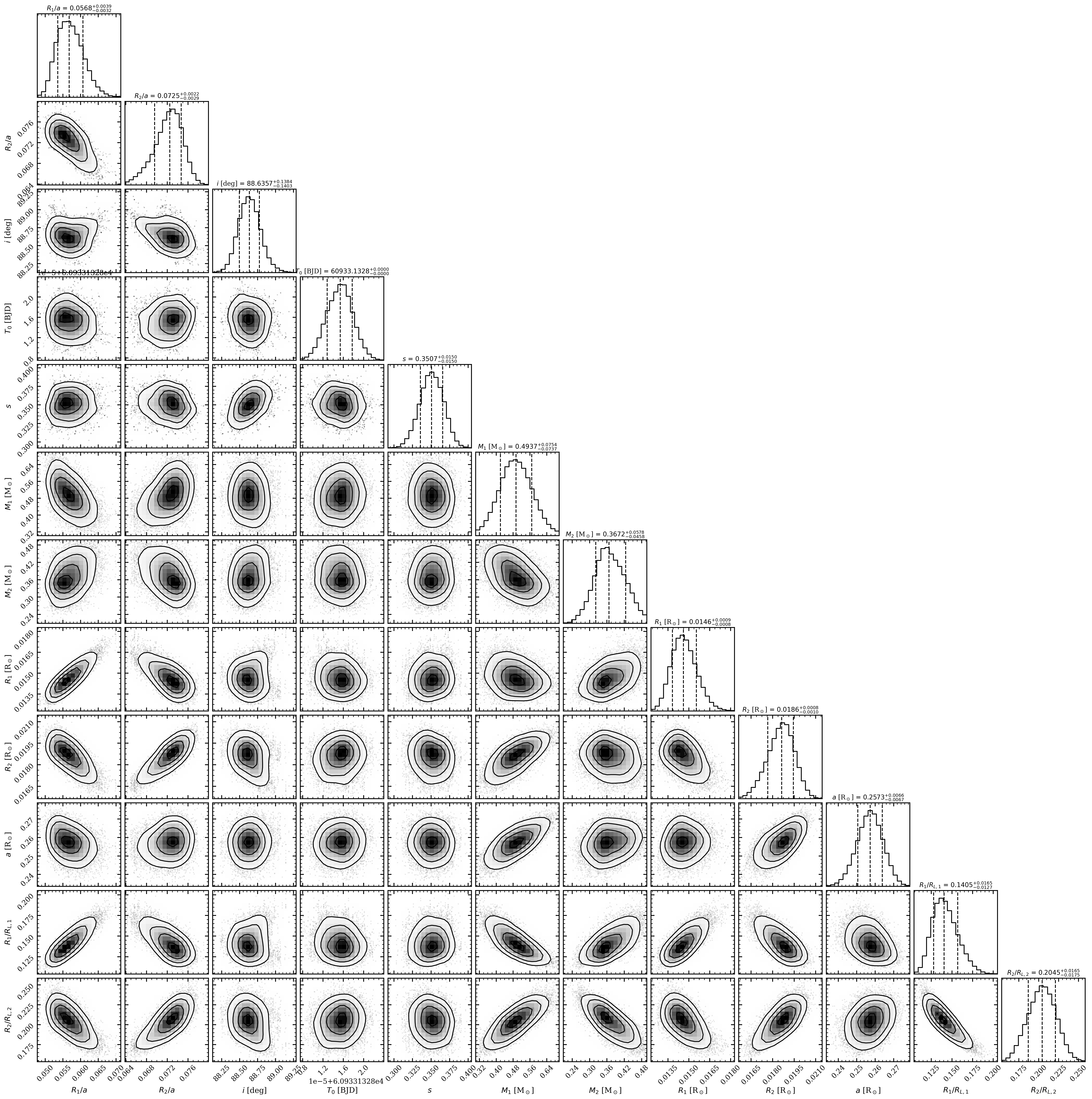}
    \caption{Posterior distributions from our \texttt{ellc} light curve modeling of ZTF\,J2229+1928. The five sampled parameters are the fractional radii of the two components $R_1/a$ and $R_2/a$ (each in units of the orbital separation $a$), the orbital inclination $i$ in degrees, the mid-eclipse time $T_0$, and the surface brightness ratio $s = J_2/J_1$. The linear limb-darkening coefficients were held fixed and are not shown. The remaining parameters are derived from each posterior sample using the carbon-oxygen core, thick-hydrogen cooling sequences of \citet{Bedard2020} for WD 1 and the helium-core cooling tracks of \citet{Althaus2013} for WD 2, evaluated at temperatures drawn with 300\,K scatter and with a 10 percent radius model error \citep{Burdge2020b}, together with Kepler's third law and the \citet{Eggleton1983} Roche lobe approximation (i.e., the component masses $M_1$ and $M_2$, the component radii $R_1$ and $R_2$ and the orbital separation $a$ in solar radii, and the Roche lobe filling factors $R_1/R_{L,1}$ and $R_2/R_{L,2}$).}

    \label{fig:placeholder}
\end{figure*}

\section{Modeling}\label{sec:modeling}

\subsection{Light Curve Modeling}
We modeled the proto-Lightspeed light curve using \texttt{ellc} \citep{Maxted2016}. The orbital period was fixed to the value measured from ZTF, while both stars were assumed to be spherical since neither star is close to filling its Roche lobe. We did not include Doppler beaming, irradiation, or gravity darkening, and the spherical model does not include ellipsoidal modulation, because the expected signals are too small to detect in these data.  For instance, irradiation is expected to be the largest of these effects for the properties of our white dwarfs, but it is still too small to detect: assuming the temperatures of 11,100\,K and 8,200\,K measured in Section \ref{sec:spectroscopy} and that all of the light from the hotter white dwarf incident on its companion is reprocessed, the irradiation amplitude is only $\sim$1,100 ppm. We adopted a linear limb-darkening law with coefficients from the DA white dwarf tables of \citet{Claret2020}, averaging the Sloan $g'$ and $r'$ values at each star's LRIS temperature and surface gravity because the proto-Lightspeed data are unfiltered \citep{Layden2026}. We fixed the coefficients to $u_1=0.362$ for the hotter white dwarf (WD 1) and $u_2=0.479$ for the cooler white dwarf (WD 2). 

The fitted parameters of our \texttt{ellc} fit were $R_1/a$, $R_2/a$, the orbital inclination $i$, the primary eclipse time $T_0$, and the surface brightness ratio $J_2/J_1$. The posterior distributions were sampled using the \texttt{emcee} Markov chain Monte Carlo sampler \citep{Foreman-Mackey2013} with 32 walkers and 5000 steps. We started the walkers from a tight distribution around a best-fitting preliminary parameter estimate, discarded the first 1000 steps as burn-in, and assessed convergence with the Gelman-Rubin statistic \citep{Gelman1992}. Because the light curve constrains only the fractional radii and inclination, a mass-radius relation for each star is needed to determine the absolute radii and individual masses. For the hotter and higher mass white dwarf, we used the carbon-oxygen core, thick hydrogen layer cooling sequences of \citet{Bedard2020}, whose grid covers 0.2 to 1.3\,$M_\odot$. For the cooler white dwarf, we used the helium-core cooling tracks of \citet{Althaus2013}, because white dwarfs below about 0.45\,$M_\odot$ are expected to have helium cores. These tracks cover 0.155 to 0.435\,$M_\odot$, so above 0.435\,$M_\odot$ we extrapolated linearly from the two heaviest tracks. For each posterior sample, we drew each LRIS temperature with 300\,K Gaussian scatter and used the cooling models to relate radius to mass and temperature. We also applied a 10 percent fractional radius model error draw, following the value adopted for white dwarf mass-radius relations by \citet{Burdge2020b}. The masses were then solved so that the model radii matched $(R_1/a)\,a$ and $(R_2/a)\,a$, with $a$ set by Kepler's third law for the total mass and fixed period. Roche lobe filling factors were calculated using the \citet{Eggleton1983} approximation. The posteriors are shown in Figure \ref{fig:placeholder} and Table \ref{tab:params}.
Importantly, we note that the quoted uncertainties in Table \ref{tab:params} are the statistical posteriors conditional on the adopted mass-radius relations and do not include systematic uncertainty from the choice of cooling model. 

\begin{table}
\centering
\caption{Parameters for ZTF\,J2229+1928. WD 1 is the hotter white dwarf and WD 2 is the cooler one. The fitted \texttt{ellc} parameters are the fractional radii $R_1/a$ and $R_2/a$ in units of the orbital separation, inclination $i$, mid-eclipse time $T_0$, and surface brightness ratio $J_2/J_1$. The Roche lobe filling factors $R_1/R_{L,1}$ and $R_2/R_{L,2}$ give the ratio of each white dwarf radius to its Roche lobe radius, calculated using \citet{Eggleton1983}. The masses, radii, and orbital separation are derived using mass-radius relations.}
\label{tab:params}
\begin{tabular}{lc}
\hline\hline
  Parameter & Value \\
\hline
R.A. (J2000) & 22:29:16.21 $\pm$ 0.01$''$ \\
Decl. (J2000) & +19:28:21.9 $\pm$ 0.01$''$ \\
Pan-STARRS ID & 131363373176327976 \\
Pan-STARRS $g$ (mag) & $21.21 \pm 0.05$ \\
Pan-STARRS $r$ (mag) & $21.14 \pm 0.04$ \\
Period (minutes) & $23.413090(13)$ \\
\hline
  $R_1/a$ & $0.057^{+0.004}_{-0.003}$ \\
  $R_2/a$ & $0.072^{+0.002}_{-0.003}$ \\
  $i$ (deg) & $88.64^{+0.14}_{-0.14}$ \\
  $T_0$ (BJD $-$ 2400000.5) & $60933.132815(3)$ \\
  $J_2/J_1$ & $0.351^{+0.015}_{-0.015}$ \\
\hline
  $M_1$ (M$_\odot$) & $0.49^{+0.08}_{-0.07}$ \\
  $M_2$ (M$_\odot$) & $0.37^{+0.06}_{-0.05}$ \\
  $R_1$ (R$_\odot$) & $0.0146^{+0.0009}_{-0.0008}$ \\
  $R_2$ (R$_\odot$) & $0.0186^{+0.0010}_{-0.0008}$ \\
  $a$ (R$_\odot$) & $0.257^{+0.007}_{-0.007}$ \\
  $R_1/R_{L,1}$ & $0.140^{+0.016}_{-0.013}$ \\
  $R_2/R_{L,2}$ & $0.204^{+0.016}_{-0.017}$ \\
\hline
\end{tabular}
\end{table}

\begin{figure*}[!t]
    \centering
    \includegraphics[width=0.95\linewidth]{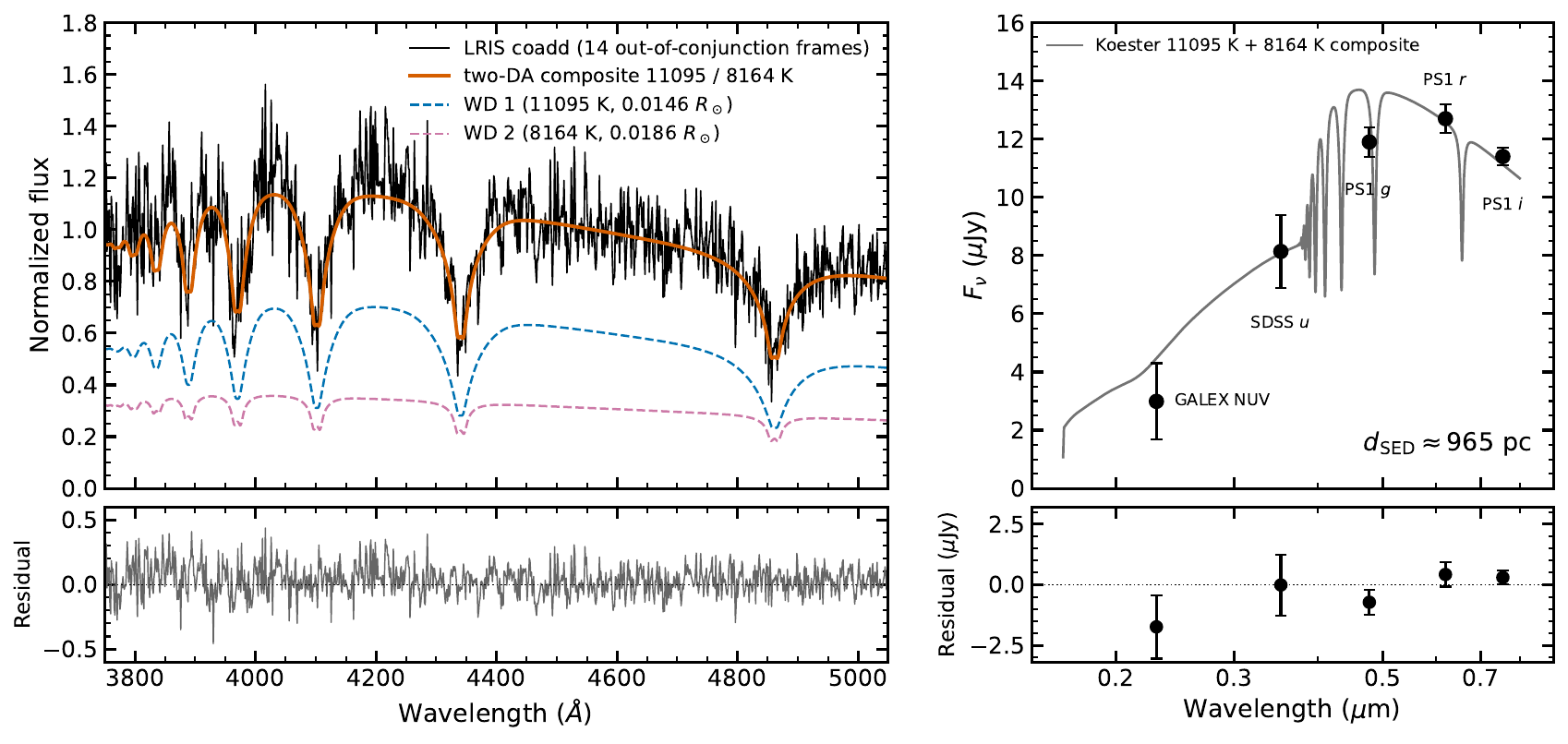}
    \caption{LRIS spectroscopy and spectral energy distribution (SED) of ZTF J2229+1928 with the best-fitting two-component model. Left) a deep stack of the 14 out-of-conjunction LRIS spectra (black) with the composite model (orange), white dwarf 1 (WD 1, blue dashed), and white dwarf 2 (WD 2, pink dashed), with the plotted models averaged over the same 14 orbital phases. Right) archival SED photometry (black points, each band labeled) with the composite LRIS fit scaled to the photometry (gray) and the implied distance. Residuals are shown beneath each panel.}
    \label{fig:sed}
\end{figure*}

\subsection{Keck/LRIS Spectroscopy}\label{sec:spectroscopy}
We obtained phase-resolved spectroscopy of ZTF J2229+1928 with the Low Resolution Imaging Spectrometer (LRIS; \citealt{Oke1995}) on the 10-m Keck I telescope on UT 2026 July 7. We acquired 22 blue-arm spectra with the 600/4000 grism and 22 red-arm spectra with the 600/7500 grating (120\,s exposures, 1.0\,arcsec longslit, D560 dichroic), covering approximately 2.4 cycles of the $P \approx 23.41$ minute orbit. The spectra were reduced and flux calibrated with the automated LRIS pipeline \texttt{lpipe} \citep{Perley2019}. The blue arm covers the Balmer series from H$\beta$ through H8, with DA Balmer absorption and no emission lines, but the components are not resolved into separate line sets and yield no radial velocities.

We built a deep stack of 14 blue-arm spectra away from the two conjunctions, interpolating each frame onto a common wavelength grid and coadding them with inverse-variance weighting. To measure the component temperatures, we fit this stack with a composite of two DA atmosphere models from \citet{Koester2010}, the two-component modeling of \citet{Burdge2020b} and the grid-based chi-squared matching of atmosphere models to an LRIS spectrum in \citet{Householder2026}. The model grid is spaced by 250\,K in effective temperature and 0.25\,dex in surface gravity, so we interpolated linearly in both to make sure that the fit was not restricted to the grid points. For each pair of trial temperatures we built the composite model from the two interpolated components. We convolved each component to the instrumental resolution of $R \approx 1000$, broadened it by the rotational velocity of a tidally locked star of that radius, and weighted it by the square of its radius, which sets the relative contribution of each star to the combined light. We then Doppler shifted each component to the mid-exposure orbital phase of every contributing frame, using radial-velocity semi-amplitudes and surface gravities fixed at the median values implied by the light curve modeling. Next, we averaged the shifted models over the same 14 frames that were included in the stack, so the model carried the same orbital smearing as the data. We also normalized the local continuum in each Balmer line window with an iterative linear fit that rejects the absorption cores, following a similar approach to \citet{Householder2026b}. We then applied the identical normalization to the model and evaluated the fit on the normalized spectra, so the temperatures were constrained by the line shapes rather than the continuum calibration. Finally, we fit the spectra with a Gaussian prior from the light curve surface brightness ratio, which helps separate the temperatures of the two stars, and measured $T_1 \approx 11{,}100$\,K and $T_2 \approx 8{,}200$\,K (Figure \ref{fig:sed}). The uncertainties on these temperatures are dominated by choices made in the fit rather than by photon noise, and are difficult to quantify, so rather than report uncertainties we simply regard both temperatures as accurate to within a few hundred kelvin. We also attempted to measure radial velocities from the individual spectra, but no coherent velocity variations were recovered. In particular, no S-wave from either component is apparent in the trailed spectrogram (Appendix \ref{app:trail}). Measuring the radial velocities of this source will thus likely require more telescope time on a 10-m class telescope, or extremely large telescopes.

\subsection{Estimating the distance to ZTF J2229+1928}
ZTF J2229+1928 does not have a measured \textit{Gaia} parallax, so its distance cannot currently be determined from astrometry. We therefore estimated the distance by fitting its SED using photometry retrieved through VizieR \citep{Ochsenbein2000}, dereddened with $E(B-V) = 0.043$ from the \citet{Schlegel1998} dust map with the \citet{Schlafly2011} recalibration and the \citet{Cardelli1989} extinction law with $R_V = 3.1$. We adopted the GALEX NUV, SDSS $u$, and Pan-STARRS $gri$ measurements, and excluded points with only upper limits or large fractional uncertainties (Figure \ref{fig:sed}). Because this is a binary system, the observed SED was fit by the sum of two white dwarf spectra:
\begin{equation}
F_\nu(\lambda) = \left(\frac{R_1}{d}\right)^2 F_{\nu,1}(\lambda) + \left(\frac{R_2}{d}\right)^2 F_{\nu,2}(\lambda),
\end{equation}
where $F_{\nu,1}(\lambda)$ and $F_{\nu,2}(\lambda)$ are the surface fluxes from hydrogen-atmosphere (DA) white dwarf models from \citet{Koester2010}, $R_1$ and $R_2$ are the component radii from the light curve modeling, and $d$ is the distance to the system. The surface gravities for each component were computed from the MCMC posteriors, and the models were interpolated to those values. With the radii and the temperatures held at their fitted values, the distance is the only free parameter. We fit the dereddened GALEX NUV, SDSS $u$, and Pan-STARRS $g$, $r$, and $i$ fluxes to the two-component model by minimizing the chi-squared ($\chi^2$) statistic, with each data point weighted by one over the square of its photometric uncertainty. From this fit, we obtained $d \approx 965$\,pc.

\section{Discussion}\label{sec:discussion}

\subsection{Gravitational-Wave Prospects}

\begin{figure}[ht]
    \centering
    \includegraphics[width=\linewidth]{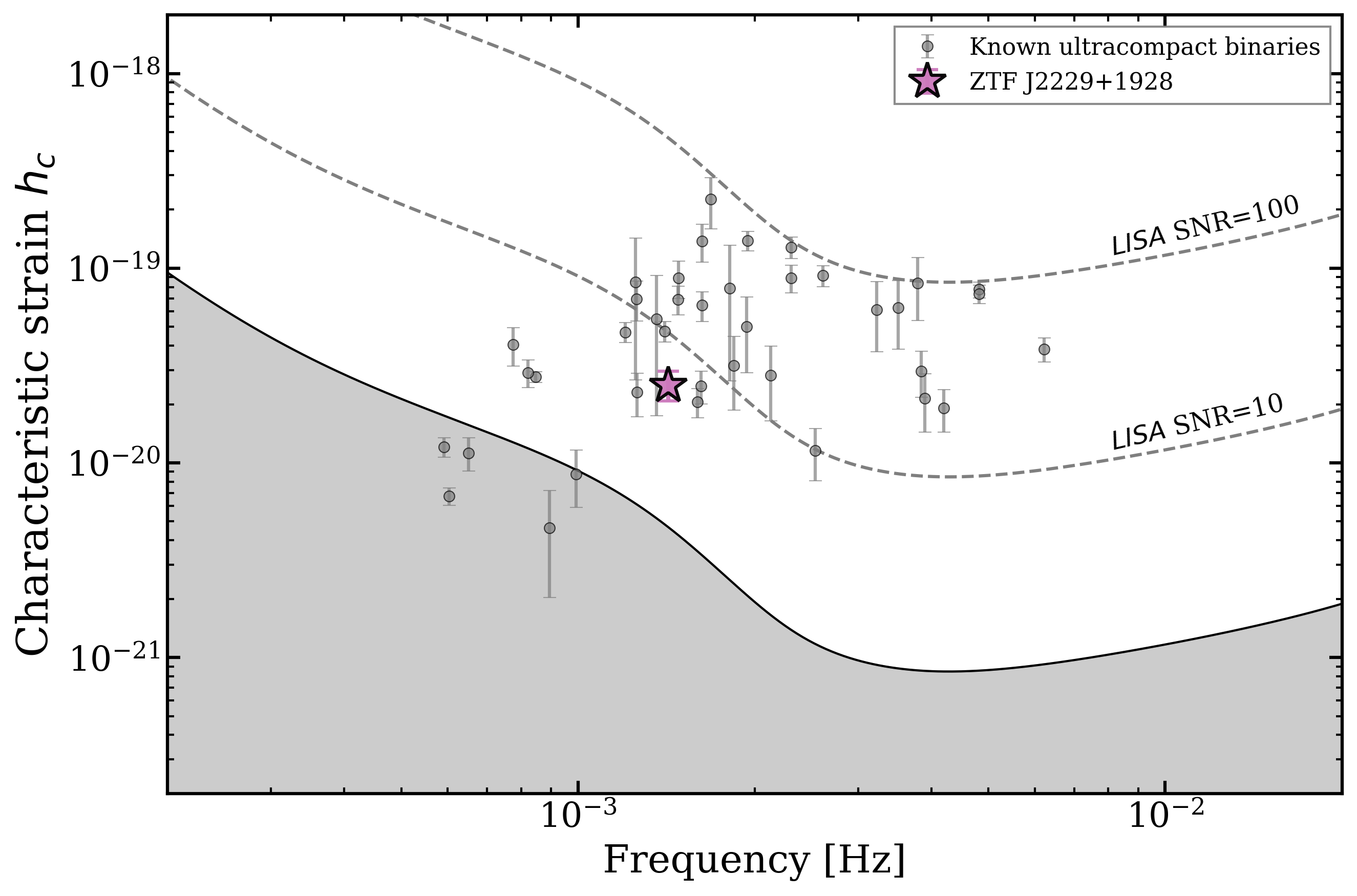}
    \caption{Predicted characteristic strain of ZTF J2229+1928 (pink). Other known ultracompact binaries compiled by \citet{Chickles2025} are plotted in gray. The two gray dashed curves mark the characteristic strain at which a monochromatic source reaches a signal-to-noise ratio of 10 and 100 over a four-year \textit{LISA} mission. We computed these curves from the \citet{Robson2019} noise model, including Galactic confusion noise, using $h_c = \mathrm{SNR}\sqrt{f S_n(f)}$.}
    \label{fig:lisa}
\end{figure}

Compact double white dwarfs like ZTF J2229+1928 are among the strongest persistent gravitational-wave sources in the Galaxy. Their small orbital separations and high orbital frequencies place them squarely in the millihertz regime where the upcoming \textit{LISA} mission is most sensitive. We estimated the expected \textit{LISA} signal-to-noise ratio (SNR) for ZTF J2229+1928 using Monte Carlo sampling over the posterior distributions of the system parameters derived in Section \ref{sec:modeling} ($M_1$, $M_2$, $i$) at a distance of $d = 965$\,pc from the SED fit. For each posterior draw, we computed the chirp mass
\begin{equation}
\mathcal{M}_c = \frac{(M_1 M_2)^{3/5}}{(M_1 + M_2)^{1/5}}
\end{equation}
and the frequency-domain gravitational-wave amplitude
\begin{equation}
    \mathcal{A} = \sqrt{\frac{5}{24}}\,\frac{1}{\pi^{2/3}}\,\frac{(G\mathcal{M}_c)^{5/6}}{c^{3/2}\,d}\,f_{\mathrm{GW}}^{-7/6},
\end{equation}
where $G$ is the gravitational constant, $c$ is the speed of light, $d$ is the distance, and $f_{\mathrm{GW}}$ is the gravitational-wave frequency. We then averaged over polarization angle and the \textit{LISA} antenna pattern to obtain the effective signal power, and compared it against the analytic \textit{LISA} noise model of \citet{Robson2019}, assuming a four-year observation baseline. At a gravitational-wave frequency of $f_{\mathrm{GW}} = 2/P = 1.42\times10^{-3}$\,Hz, we find a characteristic strain of $h_c = 2.5\times10^{-20}$, corresponding to a four-year $\mathrm{SNR} = 5.3^{+1.0}_{-0.9}$ (Figure \ref{fig:lisa}). The predicted four-year SNR falls within the $\mathrm{SNR} \approx 5$--6 range commonly adopted for individually detectable \textit{LISA} sources \citep{Kupfer2018,Finch2023}. Because the SNR grows as the square root of the observing time, a longer mission would increase the expected SNR.

\subsection{Orbital Decay and Evolutionary Fate}

For a circular orbit, gravitational radiation alone predicts $\dot{P} \approx -4.0 \times 10^{-12}$\,s\,s$^{-1}$ \citep{Peters1964} for the properties of ZTF J2229+1928. The corresponding decay timescale, $(3/8)\,P/|\dot{P}|$, is roughly 4.2\,Myr, so ZTF J2229+1928 reaches contact well within a Hubble time. What remains uncertain is whether the system survives the onset of mass transfer as a stable accreting binary or instead merges. 
 At Roche-lobe contact, material leaving the donor through the inner Lagrangian point follows a ballistic path whose closest approach to the accretor's center is about 0.0029\,$R_\odot$, assuming the component masses and radii remain fixed \citep{Nelemans2001}. Because this distance is smaller than the 0.0146\,$R_\odot$ radius of the 0.49\,$M_\odot$ white dwarf, the stream strikes the accretor instead of orbiting it to form a disk, so mass transfer is expected to begin through direct impact \citep{Marsh2004}. If mass transfer through this direct-impact phase remains stable, the orbit may widen as mass moves to the accretor, and once the stream no longer strikes the accretor a disk may form \citep{Marsh2004}. If it does not remain stable, the donor is disrupted and the binary merges. In the models of \citet{Marsh2004} with a 10\,yr synchronization timescale, binaries with accretors near 0.49\,$M_\odot$ are stable only for donors below about 0.24\,$M_\odot$. In the models of \citet{Kremer2017} with a 10\,yr synchronization timescale for both stars at contact, binaries with accretors near 0.49\,$M_\odot$ remain stable for donors up to about 0.32\,$M_\odot$. The median masses of $0.49 + 0.37\,M_\odot$ ($q \approx 0.76$) therefore give unstable mass transfer in both models, while the lower $1\sigma$ donor mass of 0.32\,$M_\odot$ reaches the stable region of \citet{Kremer2017}. Furthermore, even a binary that is considered stable in these models may still merge. For the $0.2 + 0.5$ and $0.2 + 1.0\,M_\odot$ binaries modeled by \citet{Shen2015}, drag during the first nova drives mergers within hours, leading \citet{Shen2015} to argue that most, if not all, interacting double white dwarfs likely merge. In the merger simulations of \citet{Dan2012}, a $0.49 + 0.37\,M_\odot$ binary lies outside the region that produces a helium detonation, so a Type Ia supernova is unlikely. Such mergers may produce R Coronae Borealis stars or hot subdwarfs \citep{Schwab2019,Zhang2012}, with the subsequent evolution onto the white dwarf cooling track described by \citet{Schwab2019}.

\section{Conclusion}\label{sec:conclusion}
In this paper, we have presented the discovery and characterization of ZTF J2229+1928, a detached and eclipsing double white dwarf in a 23.4 minute orbit. Our main results are summarized below:

\begin{itemize}
    \item ZTF J2229+1928 was identified in a GPU-accelerated search of ZTF photometry and confirmed with high-speed follow-up observations using proto-Lightspeed on Magellan.

    \item The high-speed light curve constrains the fractional radii and orbital inclination. Combining these constraints with the finite-temperature cooling models of \citet{Bedard2020} for WD 1 and \citet{Althaus2013} for WD 2 yields the component masses and absolute radii in Table \ref{tab:params}.

    \item Fitting the Keck/LRIS spectra with a Gaussian prior on the light curve surface brightness ratio gives effective temperatures of $T_1 \approx 11{,}100$\,K and $T_2 \approx 8{,}200$\,K, and scaling the two-component model to the broadband SED gives a distance of $\approx 965$\,pc.

    \item At a gravitational-wave frequency of $f_{\mathrm{GW}} = 1.42\times10^{-3}$\,Hz, we predict a characteristic strain of $h_c = 2.5\times10^{-20}$ and a four-year $\mathrm{SNR} = 5.3^{+1.0}_{-0.9}$, making this system a promising target for the upcoming \textit{LISA} mission.
\end{itemize}

Future follow-up will be crucial for characterizing this system. We are continuing high-speed photometric monitoring to measure the orbital period derivative, which traces the rate of orbital decay from gravitational-wave emission. Additional spectroscopic observations with large-aperture facilities could also help further characterize the properties of the DWDs, though the faintness of this system will likely make such efforts time-intensive. 

\facilities{PO:1.2m (ZTF), Magellan:Clay (proto-Lightspeed), Keck:I (LRIS)}
\software{astropy \citep{2022ApJ...935..167A}, numpy \citep{2020Natur.585..357H}, scipy \citep{2020NaMet..17..261V}, matplotlib \citep{2007CSE.....9...90H}, ellc \citep{Maxted2016}, emcee \citep{Foreman-Mackey2013}, LPipe \citep{Perley2019}}

\section*{Acknowledgments} This work was supported by the MIT Undergraduate Research Opportunities Program (UROP). This material is also based upon work supported by the National Science Foundation Graduate Research Fellowship under Grant No. 2141064 and the MIT Dean of Science Fellowship. We also acknowledge the use of computational resources provided by the MIT Engaging cluster at the Massachusetts Green High Performance Computing Center (MGHPCC).

This paper includes data gathered with the 6.5 meter Magellan Telescopes located at Las Campanas Observatory, Chile. This work was supported by a NASA Keck PI Data Award, administered by the NASA Exoplanet Science Institute. Data presented herein were obtained at the W. M. Keck Observatory from telescope time allocated to the National Aeronautics and Space Administration through the agency's scientific partnership with the California Institute of Technology and the University of California. The Observatory was made possible by the generous financial support of the W. M. Keck Foundation.

The authors wish to recognize and acknowledge the very significant cultural role and reverence that the summit of Maunakea has always had within the Native Hawaiian community. We are most fortunate to have the opportunity to conduct observations from this mountain.

This work is based on observations obtained with the Samuel Oschin Telescope 48-inch and the 60-inch Telescope at the Palomar Observatory as part of the Zwicky Transient Facility project. ZTF is supported by the National Science Foundation under Grants No. AST-1440341 and AST-2034437 and a collaboration including current partners Caltech, IPAC, the Weizmann Institute of Science, the Oskar Klein Center at Stockholm University, the University of Maryland, Deutsches Elektronen-Synchrotron and Humboldt University, the TANGO Consortium of Taiwan, the University of Wisconsin at Milwaukee, Trinity College Dublin, Lawrence Livermore National Laboratories, IN2P3, University of Warwick, Ruhr University Bochum, Northwestern University and former partners the University of Washington, Los Alamos National Laboratories, and Lawrence Berkeley National Laboratories. Operations are conducted by COO, IPAC, and UW.

\bibliography{sample7}{}
\bibliographystyle{aasjournal}

\appendix

\section{Trailed Spectrogram}\label{app:trail}

\begin{figure}[ht]
    \centering
    \includegraphics[width=\linewidth]{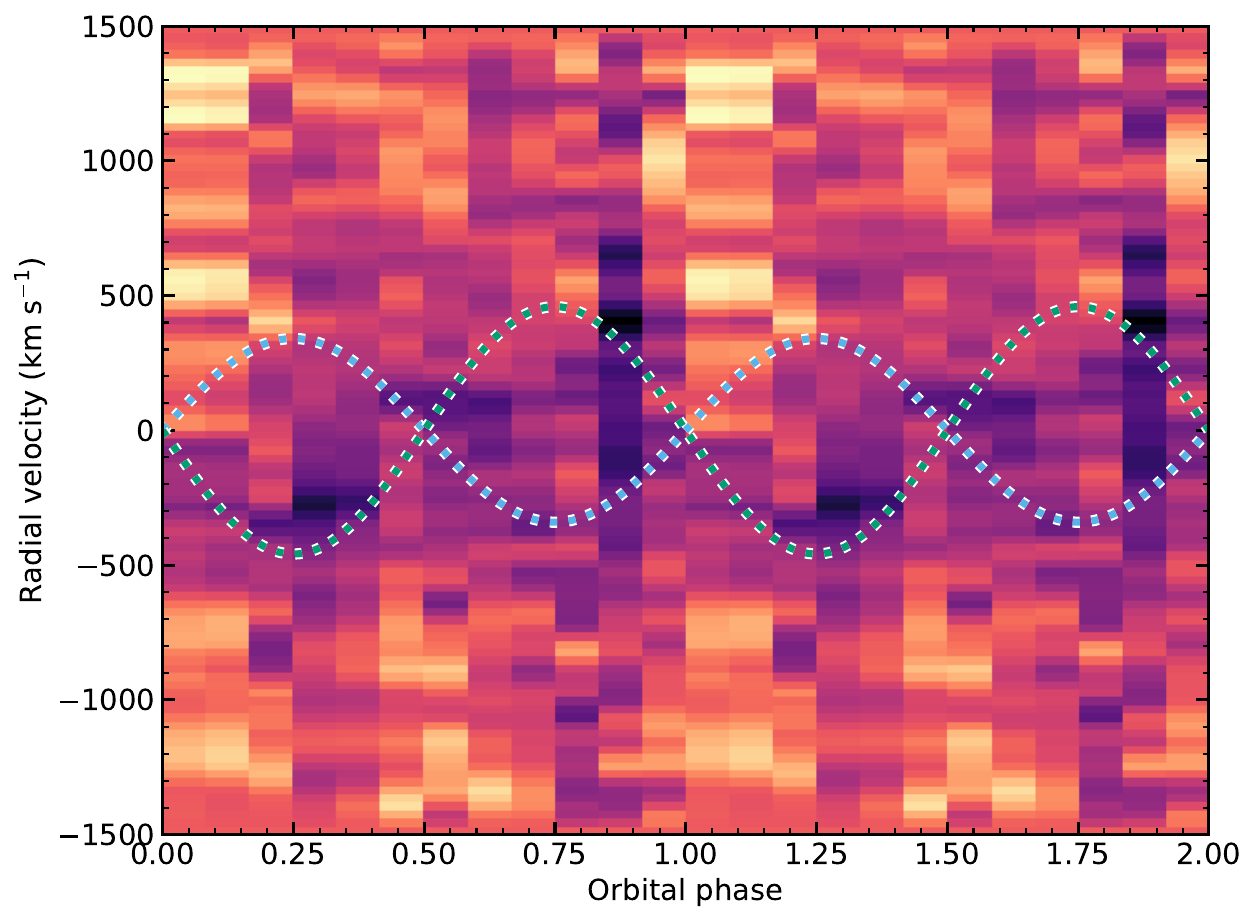}
    \caption{Trailed spectrogram of the 22 LRIS blue-arm spectra, coadding the H$\beta$ through H8 lines in velocity space and repeated over two orbital cycles. Darker shades are deeper absorption. The dotted curves show the radial-velocity curves of WD 1 (blue, $K_1 = 341\,\mathrm{km\,s^{-1}}$) and WD 2 (green, $K_2 = 459\,\mathrm{km\,s^{-1}}$) predicted from the light curve masses. No coherent S-wave from either component is detected, likely because of the low signal-to-noise ratio of the individual spectra.}
    \label{fig:trail}
\end{figure}

Figure \ref{fig:trail} shows a trailed spectrogram of the 22 LRIS blue-arm spectra, built by interpolating the five blue Balmer lines (H$\beta$ through H8) of each frame onto a common velocity grid and coadding them in 12 orbital phase bins. The radial-velocity curves of both components predicted from the light curve masses are overlaid. No coherent S-wave from either component is apparent at the signal-to-noise ratio of these data.

\end{document}